\documentclass[runningheads]{llncs}
\usepackage{graphicx}
\usepackage{amsmath}
\usepackage{amssymb}
\usepackage{tikz}
\usepackage{algorithm}
\usepackage{algpseudocode}

\newcommand\A{\mathcal{A}}
\newcommand\C{\mathcal{C}}
\newcommand\D{\mathcal{D}}
\newcommand\E{\mathcal{E}}
\newcommand\M{\mathcal{M}}
\newcommand\PP{\mathcal{P}}
\newcommand\map[3]{#1 : #2 \longrightarrow #3}
\newcommand\diagr[1]{\textbf{Diag(}#1\textbf{)}}
\newcommand\TS{\textbf{TS(L)}}
\newcommand\exec{\textbf{Ex}}
\newcommand\zig{\textbf{Zig}}

\begin{document}
\title{Bisimilarity of Diagrams\thanks{The author was supported by ERATO HASUO
Metamathematics for Systems Design Project (No.~JPMJER1603), JST and
Grant-in-aid No.~19K20215, JSPS.}}
%
%
\author{J\'er\'emy Dubut\inst{1,2}}
\authorrunning{J. Dubut}
%
\institute{National Institute of Informatics, Tokyo, Japan \and
Japanese-French Laboratory of informatics, CNRS IRL 3527, Tokyo, Japan
\email{dubut@nii.ac.jp}}
\maketitle              
\begin{abstract}
In this paper, we investigate diagrams, namely functors from any small category 
to a fixed category, and more particularly, their bisimilarity. Initially defined using 
the theory of open maps of Joyal et al., we prove two characterisations of this 
bisimilarity: it is equivalent to the existence of a bisimulation-like relation and 
has a logical characterisation \`a la Hennessy and Milner. We then prove that we 
capture both path bisimilarity and strong path bisimilarity of any small open 
maps situation. We then look at the particular case of finitary diagrams with 
values in real or rational vector spaces. We prove that checking bisimilarity and 
satisfiability of a positive formula by a diagram are both decidable by reducing 
to a problem of existence of invertible matrices with linear conditions, which in 
turn reduces to the existential theory of the reals.

\end{abstract}
\section{Introduction}

Diagrams in a category, namely functors from any small category to this specified category, are essential objects in category theory. Numerous basic constructions in category theory can be seen as a limit or colimit of a suitable diagram. However, their usefulness is not limited to those.

In the context of directed algebraic topology (see \cite{fajstrup16} for a textbook), Dubut et al. used diagrams with values in a category of modules to encode local geometric properties of a directed space and their evolution \cite{dubut15}. The domains of those diagrams are given by directed paths of the space and their extensions, while the diagrams themselves map such a path to some homology modules describing the default of directed homotopy of the space. It was then observed that a suitable notion of bisimilarity of diagrams, using the general theory of open maps from \cite{joyal96} was the right notion to compare such defaults of dihomotopy. 

In the first part of this paper, we propose to look at the general theory of bisimilarity of diagrams, extending it to any category of observations. After describing the original definition using open morphisms (Section~\ref{sec:opendiag}), we describe two equivalent characterisations. First (Section~\ref{sec:bisimdiag}), it is equivalent to the existence of a relation, similar to history preserving bisimulations of event structures from \cite{rabinovitch88}. This result generalises a result from \cite{dubut15}. Secondly (Section~\ref{sec:logic}), it has a logical characterisation, similar to a Hennessy-Milner theorem: two diagrams are bisimilar if and only if they both satisfy the same formulae of a path logic. We finally prove in Section~\ref{sec:rela} that we capture path and strong path bisimilarities of any open map situation \cite{joyal96} as the bisimilarity of a suitable notion of diagrams.

In a second part, we consider two decision problems for a class of diagrams with values in real or rational vector spaces, used in \cite{dubut15} for describing defaults of dihomotopy of geometric models of truly concurrent systems (See Section~\ref{sec:interlude}). For those diagrams, we prove in Section~\ref{sec:algo} that bisimilarity and the satisfaction of a positive formula are both decidable by reduction to a problem of invertible matrices, itself reduced to the existential theory of the reals (Section~\ref{sec:exithe}).

\textbf{Existing work:} This theory of bisimilarity of diagrams is intimately related to categorical theories of bisimulations. If the relation with open maps is developed in Sections~\ref{sec:opendiag} and \ref{sec:rela}, its relation with coalgebra is less clear. Relations between open maps and coalgebra are investigated in \cite{lasota02,wissmann19}, however those cannot be applied in general to a category of diagrams. The main problem is that diagrams are not naturally coalgebras in general, and so there is no clear relationship between open maps as described in Section~\ref{sec:opendiag}, and coalgebra homomorphisms (also called coverings of processes in \cite{winter08}). Another important related line of work is the theory of bisimilarity of presheaves \cite{cattani97}, which considers similar objects (presheaves are particular cases of diagrams), but from a very different point of view.

\section{Bisimilarity and bisimulations of diagrams}
\label{sec:bisim}

Diagrams with values in a fixed category $\A$ are functors $\map{F}{\C}{\A}$ 
from any small category to $\A$. If $\A$ is thought as a category of ``observations'' 
and $\C$ as the category of executions of a system, a diagram encodes the 
trace of observations along every execution (typically, a label), and its actions on 
morphisms of $\C$ encodes how these observations change when the system 
evolves. 


In this section, we describe the 
original form of the bisimilarity from \cite{dubut15}, defined as the existence of a 
span of particular morphisms of diagrams having some lifting properties. We 
then develop an equivalent characterization using relations, similar to 
bisimulations of event structures as introduced in \cite{rabinovitch88}.

	\subsection{Category of diagrams}
	
	The original definition of bisimilarity of diagrams was designed using particular morphisms of diagrams. Such a morphism, say from the diagram $\map{F}{\C}{\A}$ to the diagram $\map{G}{\D}{\A}$ is a pair $(\Phi,\sigma)$ with $\map{\Phi}{\C}{\D}$ a functor and $\sigma$ is a natural \emph{isomorphism} from $F$ to $G\circ\Phi$. The composition $(\Psi,\tau)\circ(\Phi,\sigma)$ is defined as $(\Psi\circ\Phi, (\tau_{\Phi(c)}\circ\sigma_c)_{c \text{ object of } \C})$. We denote this category by $\diagr{\A}$.
	
	\begin{example}
	Throughout the next two sections, we will develop a particular example of diagrams in which transition systems can be encoded. This example will allow us to relate constructions in diagrams to classical constructions in concurrency theory. From now, we fix a set $L$ called the \textbf{alphabet}. Such a set induces a poset (which can be seen as a category) $\A_L$ whose elements are the finite words on $L$ and whose order is the prefix order.
	A transition system $T$ on $L$ produces a diagram $\map{F_T}{\C_T}{\A_L}$ as follows. The category $\C_T$ is formed by considering as objects the runs of $T$, that is, sequences $i \xrightarrow{a_1} q_1 \xrightarrow{a_2} \ldots \xrightarrow{a_n} q_n$ of transitions of $T$ where $i$ is the initial state, and by ordering them by prefix. $F_T$ then maps a run to its sequence of labels. This construction extends to a functor $\Pi$ from the category $\TS$ of transition systems on $L$ to the category $\diagr{\A_L}$.
	Conversely, a diagram $\map{F}{\C}{\A_L}$ produces a transition system $T$ as follows. First, such a diagram can be identified with a diagram with values in $\TS$ by identifying a word $a_1.a_2.\ldots.a_n$ with the finite linear transition system $0 \xrightarrow{a_1} 1 \xrightarrow{a_2} \ldots \xrightarrow{a_n} n$. $T$ is then obtained by forming the colimit of this diagram in $\TS$. This extends to a functor $\Gamma$ from $\diagr{\A_L}$ to $\TS$.
	Note that $\Gamma\circ\Pi$ is the unfolding of transition systems and that $\Gamma$ is the left adjoint of $\Pi$. 
	\end{example}
	
	The reason why we need natural isomorphisms in the definition of a morphism of diagram is not clear yet, as the only isomorphisms in the category $\A_L$ are the identities. This will be illustrated in the case where $\A$ is a category of vector spaces. Intuitively, two isomorphic vector spaces represent the same kind of observations (in the case of directed algebraic topology, the same kind and number of holes), which we do not want to discriminate.
	
	\subsection{Open morphisms of diagrams}
	\label{sec:opendiag}
	
	The original idea from \cite{dubut15} was to compare diagrams similarly to transition systems using the theory of \cite{joyal96}. Let us call \textbf{branch} a diagram from $\textbf{n}$ to $\A$ for $n \in \mathbb{N}$, where $\textbf{n}$ is the poset (seen as a category) $\{1,\ldots,n\}$ with the usual ordering. An \textbf{evolution} of a diagram $\map{F}{\C}{\A}$ is then a morphism from any branch to $F$. Much as transition systems and executions, a morphism of diagrams $(\Phi,\sigma)$ from $\map{F}{\C}{\A}$ to $\map{G}{\D}{\A}$ maps evolutions of $F$ to evolutions of $G$: if $(\Psi,\tau)$ is an evolution of $F$, i.e., a morphism from a branch to $F$, then $(\Phi,\sigma)\circ(\Psi,\tau)$ is an evolution of $G$. Then morphisms act as particular simulations of diagrams. The idea from \cite{joyal96} was to provide conditions on morphisms for them to act as particular bisimulations. The general idea is that a morphism induces a bisimulation if it lifts evolutions of $G$ to evolutions of $F$. In the context of diagrams, this will be defined using \textbf{extensions of branches}. An extension of a branch $\map{B}{\textbf{n}}{\A}$ is a morphism of diagrams $(\Pi,\theta)$ from $\map{B}{\textbf{n}}{\A}$ to a branch $\map{B'}{\textbf{n}'}{\A}$, with $n' \geq n$ such that:\\
		\indent -- for every $i \leq n$, $B(i) = B'(i)$,\\
		\indent -- for every $i \leq j \leq n$, the morphism $B'(i\leq j)$ of $\A$ is equal to $B(i\leq j)$,\\
		\indent -- for every $i \leq n$, $\Pi(i) = i$,\\
		\indent -- for every $i \leq n$, $\theta_i = id_{B(i)}$.
	\begin{figure}
	\begin{center}
    	\begin{tikzpicture}[scale=1]
		
	\node (1) at (0,0) {$B(1)$};
   	\node (2) at (2.5,0) {$\ldots$};
	\node (3) at (5,0) {$B(n)$};
	\node (1p) at (0,-1) {$B'(1)$};
	\node (2p) at (2.5,-1) {$\ldots$};
	\node (3p) at (5,-1) {$B'(n)$};
	\node (4p) at (7.5,-1) {$\ldots$};
   	\node (5p) at (10,-1) {$B'(n')$};
	
	\path[->,font=\tiny]
		(1) edge node[above]{$B(1\leq 2)$} (2)
		(2) edge node[above]{$B(n-1 \leq n)$} (3)
		(1) edge node[left]{id} (1p)
		(3) edge node[right]{id} (3p)
		(1p) edge node[below]{$B'(1\leq 2)$} (2p)
		(2p) edge node[below]{$B'(n-1 \leq n)$} (3p)
		(3p) edge node[below]{$B'(n \leq n+1)$} (4p)
		(4p) edge node[below]{$B'(n'-1 \leq n')$} (5p);
			
\end{tikzpicture}
\caption{Extension of branches}
	\vspace{-0.5cm}
  			\end{center}
		\end{figure}
		
		Those conditions mean that the restriction of $B'$ to $\textbf{n}$ is $B$ and that the morphism $(\Pi,\theta)$ is the inclusion of $B$ in $B'$.

	Following \cite{joyal96}, we then say that a morphism $(\Phi,\sigma)$ from $\map{F}{\C}{\A}$ to $\map{G}{\D}{\A}$ is \textbf{open} if for every diagram of the form (in plain):

	\begin{figure}
		\vspace{-0.5cm}
			\begin{center}
    				\begin{tikzpicture}[scale=1]
		
	\node (B) at (0,0) {$B$};
   	\node (F) at (2,0) {$F$};
	\node (B') at (0,-1) {$B'$};
	\node (G) at (2,-1) {$G$};
	
	\path[->,font=\scriptsize]
		(B) edge node[left]{$(\Pi,\theta)$} (B')
		(F) edge node[right]{$(\Phi,\sigma)$} (G)
		(B) edge node[above]{$(\Psi,\tau)$} (F)
		(B') edge node[below]{$(\Psi',\tau')$} (G);		
		
	\path[->, dotted, font = \scriptsize]
		(B') edge node[left]{$\exists$} (F);
\end{tikzpicture}
	\vspace{-1cm}
  			\end{center}
		\end{figure}
\noindent where $(\Pi,\theta)$ is an extension of branches, there is an evolution of $F$ (in dots) which makes the two triangles commute. This means that if we can extend an evolution of $F$, mapped on an evolution of $G$ by $(\Phi,\sigma)$, as a longer evolution of $G$, then we can extend it as a longer evolution of $F$ that is mapped to this longer evolution of $G$. This means in particular that $F$ and $G$ have exactly the same evolutions. As observed in \cite{dubut15}, the definition of an open map can be simplified as follows:
	
	\begin{theorem}
	\label{simpli}
	A morphism $(\Phi,\sigma)$ is open if and only if:\\
	\indent -- $\Phi$ is surjective on objects, i.e., for every object $d$ of $\D$, there is an object $c$ of $\C$ such that $\Phi(c) = d$,\\
	\indent -- $\Phi$ is a fibration, i.e., for every morphism of $\D$ of the form $\map{j}{\Phi(c)}{d'}$, there is a morphism $\map{i}{c}{c'}$ of $\C$ such that $\Phi(i) = j$.
	\end{theorem}
	
	Following \cite{joyal96}, we say that two diagrams $\map{F}{\C}{\A}$ and $\map{G}{\D}{\A}$ are \textbf{bisimilar} if there is a span of open morphisms between them, that is, a diagram $\map{H}{\E}{\A}$ and two open morphisms, one from $H$ to $F$, one from $H$ to $G$.
	
	\begin{example}
	\label{exa:opents}
	In the case of diagrams in $\A_L$, the notion of open morphisms is related to the notion of open morphisms of transition systems as defined in \cite{joyal96}. First, an open morphism $\map{f}{T}{S}$ between transition systems always induces an open morphism $\map{\Pi(f)}{\Pi(T)}{\Pi(S)}$ between the associated diagrams. In particular, if two transition systems are bisimilar then their diagrams are bisimilar. The converse also holds but proving it using open morphisms is hard (the reason will be explained later). For example, we may expect that an open morphism of diagrams of the form $\map{\Phi}{F}{\Pi(T)}$ induces an open morphism between transition systems $\map{\Gamma(\Phi)}{\Gamma(F)}{\Gamma\circ\Pi(T)}$, but that is not true in general.
	\end{example}
	
	\subsection{Bisimulations of diagrams}
	\label{sec:bisimdiag}
	
	In this section, we generalise a notion of bisimulation relations from \cite{dubut15}, which is equivalent to the existence of a span of open morphisms. This result is an equivalent of Theorem~3.1 in \cite{winter08} in the context of open maps of diagrams.
	
	A \textbf{bisimulation} $R$ between two diagrams $\map{F}{\C}{\A}$ and $\map{G}{\D}{\A}$ is a set of triples $(c,f,d)$ where $c$ is an object of $\C$, $d$ is an object of $\D$ and $\map{f}{F(c)}{G(d)}$ is an isomorphism of $\A$ such that:\\
	\indent -- for every $(c,f,d)$ in $R$ and $\map{i}{c}{c'} \in \C$, there exist $\map{j}{d}{d'} \in \D$ and $\map{g}{F(c')}{G(d')} \in \A$ such that $g\circ F(i) = G(j) \circ f$ and $(c',g,d') \in R$,\\
	
		\begin{figure}[H]
			\begin{center}
					\vspace{-1cm}
    				\begin{tikzpicture}[scale=1]
		
	\node (x') at (0,0) {$c'$};
   	\node (x) at (0,1.5) {$c$};
	\node (Fx') at (1,0) {$F(c')$};
	\node (Fx) at (1,1.5) {$F(c)$};
	\node (Gy') at (4,0) {$G(d')$};
	\node (Gy) at (4,1.5) {$G(d)$};
	\node (y') at (5,0) {$d'$};
   	\node (y) at (5,1.5) {$d$};
	
	\path[->,font=\scriptsize]
		(x) edge node[left]{$i$} (x')
		(Fx) edge node[left]{$F(i)$} (Fx')
		(Fx) edge node[above]{$f$} (Gy);
		
	\path[->, dotted, font = \scriptsize]
		(y) edge node[right]{$j$} (y')
		(Gy) edge node[right]{$G(j)$} (Gy')
		(Fx') edge node[below]{$g$} (Gy');
			
\end{tikzpicture}
	\vspace{-1cm}
  			\end{center}
		\end{figure}
		
	\indent -- symmetrically, for every $(c,f,d)$ in $R$ and $\map{j}{d}{d'}\in \D$, there exist $\map{i}{c}{c'} \in \C$ and $\map{g}{F(c')}{G(d')} \in \A$ such that $g\circ F(i) = G(j) \circ f$ and $(c',g,d') \in R$,\\
	\indent -- for all $c\in\C$, there exists $d$ and $f$ such that $(c,f,d)\in R$,\\
	\indent -- for all $d\in\D$, there exists $c$ and $f$ such that $(c,f,d)\in R$.

\begin{theorem}
\label{bisim}
Two diagrams are bisimilar if and only if there is a bisimulation between them.
\end{theorem}

\begin{example}
In the case of diagrams in $\A_L$, a bisimulation between diagrams $\Pi(T)$ and $\Pi(S)$ is just a rephrasing for a path bisimulation in the sense of \cite{joyal96} between the transition systems $T$ and $S$. In the particular case of transition systems, the existence of a path bisimulation is equivalent to the existence of a strong path bisimulation and is equivalent to the existence of a bisimulation. Consequently:
\end{example}

\begin{proposition}
Two transition systems $T$ and $S$ are bisimilar if and only if the diagrams $\Pi(T)$ and $\Pi(S)$ are bisimilar.
\end{proposition}

\section{Diagrammatic path logic}
\label{sec:logic}

In this section, we focus on a logical characterization of bisimilarity of diagrams. The logic used, which we call \textbf{diagrammatic path logic}, is similar to the logic introduced in \cite{hennessy80} for transition systems, or to path logics developed in \cite{joyal96}. A formula in this logic allows one to express that a diagram has some kind of evolutions or not.

	
	The formulae used are generated by the following grammar:
	$$\text{\textbf{Object formulae:~~}} S::=[x]P~~~~~x\in \text{Ob}(\A)$$
$$\text{\textbf{Morphism formulae:~~}} P::=\langle f \rangle P\mid?S\mid\neg P\mid \bigwedge\limits_{i\in I} P_i~~~~~f\in \text{Mor}(\A) \text{ and } I \text{ a set}$$
\noindent where $\text{Ob}(\A)$ is the class of objects of $\A$ and $\text{Mor}(\A)$ is its class of morphisms.

Intuitively, the object formula $[x]P$ means that the current object is isomorphic to $x$, and the morphism formula $\langle f \rangle P$ means that from the current object, one can fire a transition labelled by a morphism equivalent (in the sense of matrices, or conjugate in the language of group theory) to $f$. Observe that we have arbitrary conjunctions, in particular infinite and empty (we will denote the empty conjunction by $\top$).

\begin{example}
In the case of diagrams in $\A_L$, $[w]\top$ means that the current run is labeled by the word $w$ and $\langle w \leq w' \rangle \top$ means that the current run is labeled by $w$ and that it can be extended to a run labeled by $w'$. The idea is very similar to the Hennessy-Milner logic \cite{hennessy80} and the forward path logic \cite{joyal96}. The next theorem proves that, for two transition systems, satisfying the same Hennessy-Milner formulae, forward path formulae or path formulae is the same as their diagrams satisfying the same diagrammatic formulae.
\end{example}

For a diagram $\map{F}{\C}{\A}$, an object $c$ of $\C$, and an isomorphism $f$ of $\A$ of the form $\map{f}{F(d)}{x}$ for some $d$ and $x$, we define $F,c \models S$ for an object formula $S$ and $F,f,d\models P$ for a morphism formula $P$ by induction on $S$ (resp. $P$) as follows:\\
	\indent -- $F,f,c \models \top$ always,\\
	\indent -- more generally, $F,f,c\models \bigwedge\limits_{i\in I} P_i$ iff for all $i\in I$, $F,f,c\models P_i$,\\
	\indent -- $F,c \models [x]P$ iff there exists an isomorphism $\map{f}{F(c)}{x}$ of $\A$ such that $F,f,c\models P$,\\
	\indent -- for every $\map{g}{x}{x'}$, $F,f,c\models \langle g \rangle P$ iff there exists $\map{i}{c}{c'}$ in $\C$ and an isomorphism $\map{h}{F(c')}{x'}$ such that $h\circ F(i) = g\circ f$ and $F,h,c'\models P$,
	\begin{center}
	\begin{tikzpicture}[scale=1.5]
		\node (A) at (0,0.8) {$F(c)$};
		\node (B) at (1.5,0.8) {$F(c')$};
		\node (C) at (0,0) {$x$};
		\node (D) at (1.5,0) {$x'$};
		\path[->,font=\scriptsize,dotted]
		(A) edge node[above]{$F(i)$} (B)
		(B) edge node[right]{$h$} (D);
		\path[->,font=\scriptsize]
		(C) edge node[below]{$g$} (D)
		(A) edge node[left]{$f$} (C);
	\end{tikzpicture}
	\end{center}
	
	\indent -- $F,f,c\models ?S$ iff $F,c\models S$,\\
	\indent -- $F,f,c\models \neg P$ iff $F,f,c\not\models P$.

We say that a diagram $\map{F}{\C}{\A}$ is \textbf{logically simulated} by another diagram $\map{G}{\D}{\A}$ if for every object $c$ of $\C$, there exists an object $d$ of $\D$ such that for all object formula $S$, $F,c \models S$ iff $G,d \models S$. Two diagrams $F$ and $G$ are \textbf{logically equivalent} if $F$ is logically simulated by $G$ and vice-versa.

\begin{theorem}
\label{logic}
Two diagrams are bisimilar iff they are logically equivalent.
\end{theorem}

\section{Relation to path bisimilarities of open maps situations}
\label{sec:rela}


In Section~\ref{sec:opendiag}, we understood bisimilarity of diagrams using branches, as the existence of a span of open maps. In the context of \cite{joyal96}, it means that diagrams, together with their subcategory of branches is an \textbf{open map situation}. Concretely, an open map situation is a category $\M$, called the category of \textbf{systems} (in our case $\diagr{\A}$), together with a subcategory $\PP \hookrightarrow \M$, said of \textbf{paths} (here the subcategory of branches). Another typical example is the category of transition systems $\TS$ with its subcategory of finite linear systems.

In \cite{joyal96}, two notions of bisimulations between objects of $\M$ are described: the \textbf{path bisimulations} and the \textbf{strong path bisimulations}. However, for them to make sense, some conditions on the open map situation are required: $\PP$ need to be small and $\PP$ and $\M$ must have a common initial object, which we denote by $I$. 
For example, the open maps situation of transition systems satisfies those requirements, while the one of diagrams does not in general (smallness is the issue).

Concretely, a path bisimulation $R$ between objects $X$ and $Y$ of $\M$ is a set of pairs of morphisms of the form $(\map{x}{P}{X},\map{y}{P}{Y})$ for some object $P$ of $\PP$ such that:\\
	\indent -- The pair of initial morphisms $(I\rightarrow X,I\rightarrow Y)$ belongs to $R$.\\
	\indent -- For every $(\map{x}{P}{X},\map{y}{P}{Y})$ in $R$, and for every morphisms $\map{p}{P}{Q}$ of $\PP$ and $\map{x'}{Q}{X}$ with $x'\circ p = x$, there is a morphism $\map{y'}{Q}{Y}$ such that $y'\circ p = y$ and $(x',y') \in R$.\\
	\indent -- Symmetrically, swapping the roles of $X$ and $Y$.\\
Furthermore, we say that $R$ is strong if it additionally satisfies that for every $(\map{x}{P}{X},\map{y}{P}{Y})$ in $R$, and for every morphism $\map{p}{Q}{P}$ of $\PP$, $(x\circ p,y\circ p) \in R$. 

It has to be remarked that those bisimulations induce different notions of bisimilarity in general. Furthermore, they both are different to the existence of a span of open morphisms, although strong path bisimilarity coincide with this existence in many concrete cases \cite{dubut16b}. In the case of transition systems and finite linear systems, those three notions coincide.

We propose now to characterise those two notions of bisimulations using diagrams, namely, we will now describe two functors $\map{\exec}{\M}{\diagr{\PP}}$ and $\map{\overline{\exec}}{\M}{\diagr{\overline{\PP}}}$ such that the existence of a path (resp. strong path) bisimulation between $X$ and $Y$ is equivalent to the fact that $\exec(X)$ and $\exec(Y)$ (resp. $\overline{\exec(X)}$ and $\overline{\exec(Y)}$) are bisimilar as diagrams. 

First, $\exec(X)$ is the functor from $\PP \downarrow X$, the category of morphisms from any objects of $\PP$ to $X$ and commutative triangles, to the category $\PP$ which maps any morphism $\map{x}{P}{X}$ to $P$ and every commutative triangles $x'\circ p = x$ to $p$. 

\begin{theorem}
\label{theo:pbb}
$X$ and $Y$ are path bisimilar if and only if $\exec(X)$ and $\exec(Y)$ are bisimilar.
\end{theorem}


Given a category $\C$, we denote by $\zig(\C)$ the category whose objects are those of $\C$ and whose morphisms are generated by those of $\C$ and those of $\C^{\text{op}}$. This naturally extends to an endofunctor $\map{\zig}{\textbf{Cat}}{\textbf{Cat}}$. We then define $\overline{\exec} = \zig(\exec)$ from $\zig(\exec(X))$ to $\zig(\PP)$.

\begin{theorem}
\label{theo:spbb}
$X$ and $Y$ are strong path bisimilar if and only if $\overline{\exec}(X)$ and $\overline{\exec}(Y)$ are bisimilar.
\end{theorem}

\begin{remark}
This pattern of characterising a notion bisimilarity as bisimilarity of suitable diagrams whose domain is a category of ``runs'' and whose codomain is a category of ``observations'' is more general than for (strong) path bisimilarity of open maps situations, and can be pursued for Higher-Dimensional Automata \cite{pratt91,vanglabbeek05} for example.
\end{remark}

\section{Interlude}
\label{sec:interlude}

In the first part of the paper, we focused on the general theory of bisimilarity of diagrams and its relationship with usual notions of bisimilarity of transition systems (and so of process algebra). In the second part of the paper, we would like to turn our attention to other kinds of diagrams that appeared in the theory of directed algebraic topology \cite{dubut15}. While diagrams in Section~\ref{sec:rela} were typically with values in a category of words, we will now consider diagrams with values in modules on a ring. More precisely, we will focus on the following two problems for such diagrams:\\
	\indent -- \textbf{bisimilarity:} given two diagrams, are they bisimilar?\\
	\indent -- \textbf{diagrammatic model-checking:} given a diagram $F$, an object $c$ of its domain and a state formula $S$, does $F,c \vDash S$ hold? \\
The difficulty of those problems lies in the possibility to decide whether two modules are isomorphic, problem which does not appear in the context of process algebras and transition systems. Indeed, it is known that those problems are decidable in the category of transition systems (see \cite{srba13} for a dynamic list of such (un)decidability results), while they would be undecidable for diagrams with values in groups and group morphisms because it is undecidable whether two groups are isomorphic. In this paper, we will focus on the category of finite dimensional real or rational vector spaces and matrices. 

More precisely, we will stick to finitary diagrams and finitary positive formulae defined as follows. By a \textbf{finitary diagram} $F$, we mean the following data:\\
	\indent -- a finite poset $(\C, \leq)$, the \textbf{domain},\\
	\indent -- for every element $c$ of $\C$, a natural number $F(c)$ (which stands for the real vector space $\mathbb{R}^{F(c)}$),\\
	\indent -- for every pair $c \leq c'$ of $\C$, a matrix $F(c\leq c')$ of size $F(c)\times F(c')$, with coefficients in rational numbers, presented as the list of all its elements,\\
such that:\\
	\indent -- $F(c\leq c)$ is the identity matrix,\\
	\indent -- for every triple $c \leq c' \leq c''$, $F(c\leq c'') = F(c'\leq c'').F(c\leq c')$, where `.' denotes the matrix multiplication.\\
In short, a finitary diagram is a functor from a finite poset to the category of matrices with coefficients in rational numbers. One may argue that those assumptions are not reasonable, because they are not satisfied by the diagrams from Section \ref{sec:rela} as soon as there is a loop. The reason is that when deciding this bisimilarity, there are two problems: finding out how to relate the executions and constructing the bisimulation, in particular, the isomorphism part. Loops make the first part difficult, because this relation is necessarily infinite in this case. In this paper, we want to focus on the second problem because: 1) reducing the problem of existence of a bisimulation to a problem of isomorphisms in a category is the main difference from existence of bisimulation for process algebra, 2) solving this question addresses the problem of comparing natural homologies of geometric models of true concurrency from \cite{dubut15}.

We call \textbf{finitary formulae}, the formulae generated by the following grammar:
$$\text{\textbf{Object formulae:~~}} S::=[n]P~~~~~n\in\mathbb{N}$$
$$\text{\textbf{Morphism formulae:~~}} P::=\langle M \rangle P\mid?S\mid\neg P\mid \top \mid P_1\wedge P_2$$
where $M$ is a matrix with coefficients in rational numbers.
Here, $[n]P$ stands for $[\mathbb{R}^n]P$ which makes finitary formulae diagrammatic formulae in real vector spaces. This time, since we only have finitely branching diagrams, we only consider finite conjunctions. We will more particularly consider \textbf{positive} formulae, i.e., formulae without any occurrences of the negation. For example, a formula of the form $\langle M_1 \rangle \ldots \langle M_k \rangle \top$ means that there is a sequence of matrices $N_1, \ldots, N_k$ in the diagrams where $N_i$ is equivalent to $M_i$, and those equivalences are natural (in the categorical meaning).

In this case, bisimilarity and model checking problems become a problem of existence of invertible matrices satisfying some linear conditions, as we will see in Section \ref{sec:algo}. In Section \ref{sec:exithe}, we will start by proving that this problem of matrices can be encoded in the existential theory of the reals, which is known to be decidable.

\section{Existential theory of invertible matrices}
\label{sec:exithe}

In the present section, we focus on an existential theory of matrices. We first recall the case of the existential theory of the reals, which is known to be decidable. We then introduce the existential theory of invertible matrices in $\mathbb{R}$ and $\mathbb{Q}$ and we finally prove the decidability of their satisfiability problems.

\subsection{The existential theory of some rings}

Designing algorithms for finding solutions of equations is an old problem in mathematics. The famous Hilbert's tenth problem posed the problem for polynomial equations in integers, but the question can be asked for other rings. Tarski in \cite{tarski51} solved this question for real numbers: the first-order logic of real closed fields is decidable, although the solution is of non-elementary complexity. Several improvements have been made: it was proved to be in EXPSPACE in \cite{benor86} and that the existential theory of the reals is in PSPACE in \cite{canny88}. On the contrary, Matiyasevich's negative answer of the tenth problem \cite{matiyasevitch93}, means that the existential theory of the integers is undecidable. In particular, since it is possible to express that a rational number is an integer (using possibly universal quantifiers), the full first-order logic of the rationals is undecidable. However, it is still an open question whether its existential fragment is decidable or not.

\subsection{Theory of matrices}

In this section, we will consider a logic of matrices that will be expressible in the existential theory of the reals. It will be the main ingredient to decide some problems in diagrams with values in vector spaces. Namely, we consider formulae of the form:
$$\exists_{n_1}X_1.\ldots.\exists_{n_k}X_k. \bigwedge\limits_{j = 1}^m P_j(X_1, \ldots, X_k)$$
where:\\
	\indent -- $n_i \geq 0$, is a natural number,\\
	\indent -- $X_i$ is a variable ranging over invertible matrices of dimension $n_i$,\\
	\indent -- $P_j$ is a predicate of the form $A.X_i = X_k.B$ for some $i$, $k$ and matrices $A$, $B$ with coefficients in rational numbers, $A$ and $B$ are of size $n_k\times n_i$, and $.$ denotes the matrix multiplication. We call it the \textbf{existential theory of invertible matrices}.

We will consider the following decision problem: given such a formula, is it satisfiable, that is, are there matrices $M_1$, ..., $M_k$, with $M_i$ of size $n_i\times n_i$, invertible such that for every $j$, $P_j(M_1, ..., M_k)$ is true?

We may ask this question for matrices $M_i$ in coefficients in real or rational numbers. We will prove that both problems actually coincide and are decidable in PSPACE.

\subsection{Decidability in $\mathbb{R}$}

We stick here to the case of real numbers. We prove that we have a reduction to the existential theory of the reals. Given a formula
$$\Phi = \exists_{n_1}X_1.\ldots.\exists_{n_k}X_k. \bigwedge\limits_{j = 1}^m P_j(X_1, \ldots, X_k)$$
we will construct a formula $\Psi$ in the existential theory of the reals which is satisfiable if and only if $\Phi$ is.

First, for every variable $X_i$, check if it appears in some $P_j$. If not, forget it. Indeed, if it does not appear in any predicate, then we can just choose the identity. Then, for every other quantifier $\exists_{n_i}X_i$, we fix $2.n_i^2$ fresh first-order variables $x_i^{r,s}$ and $y_i^{r,s}$ for $r,s \in \{1, ..., n_i\}$. Let $X_i$ be the matrix of size $n_i\times n_i$ whose coefficients are $x_i^{r,s}$, and $Y_i$ whose coefficients are $y_i^{r,s}$. Developing $A.X_i = X_j.B$ leads to $n_jn_i$ linear equations on the variables $x_i^{r,s}$ and $x_j^{r,s}$. So every predicate $P_j$ induces a set $L_j$ of linear equations. It remains to express that $X_i$ is invertible in the first-order logic. The idea is to express that $Y_i$ is its inverse. Developing $X_i.Y_i = \text{Id}$ and $Y_i.X_i = \text{Id}$, leads to $2.n_i^2$ polynomial equations on the variables $x_i^{r,s}$ and $y_i^{r,s}$. Let $S_i$ be the set of these equations. We denote by $\Psi$ the formula:
$$\exists x_1^{1,1}.\ldots\exists x_k^{n_k,n_k}.\exists y_1^{1,1}.\ldots\exists y_k^{n_k,n_k}. \bigwedge\limits_{i=1}^k S_i \wedge \bigwedge\limits_{j=1}^m L_j$$
$\Psi$ is of polynomial size on the size of $\Phi$: indeed, the only problem might be that we fix $2n_i^2$ variables while $n_i$ is of size $\log(n_i)$, which may say that we fix an exponential number of variables. The point is that if we fixed those $2n_i^2$ variables, then it means that $X_i$ appears in some $P_j$, and that the matrices appearing in $P_j$ have a polynomial size in $n_i$. Consequently, we fix only a polynomial number of variables.

\begin{theorem}
$\Psi$ is satisfiable in the existential theory of the reals iff $\Phi$ is satisfiable in the existential theory of invertible matrices with coefficients in real numbers. Consequently, the existential theory of invertible matrices with coefficients in real numbers is decidable in PSPACE.
\end{theorem}

\subsection{The rational case}

As we have seen previously, first-order theories of rationals are in general harder than those in reals. But there are some algebraic problems that are known to coincide when considering real and rational numbers. Given a linear system with coefficients in rational numbers, Gaussian elimination works independently of the coefficient field. Consequently, the real subspace $F_{\mathbb{R}}$ of solutions of this system has the same dimension as the rational subspace $F_{\mathbb{Q}}$ of solutions of the system. Actually, $F_{\mathbb{R}}\cap \mathbb{Q}^n = F_{\mathbb{Q}}$ and they have a common basis whose vectors are with coefficients in rational numbers. Similarly, the problem of equivalence of matrices coincides in the fields of real and rational numbers. Given two matrices $A$ and $B$ with coefficients in rational numbers, $A$ and $B$ are equivalent if there are two invertible matrices $X$ and $Y$ such that $A.X = Y.B$. This problem is also solvable using Gaussian elimination by computing the rank of $A$ and $B$, which is independent of the coefficient field. Our problem is a generalization of the equivalence problem and it is not surprising that the same kind of results hold:

\begin{theorem}
\label{rationals}
A formula $\Phi$ is satisfiable in the existential theory of invertible matrices with coefficients in real numbers if and only if it is satisfiable in the existential thoery of invertible matrices with coefficients in rational numbers.
\end{theorem}

\section{Decidability in diagrams}
\label{sec:algo}

Finally, we prove two decidability results for bisimilarity of diagrams and diagrammatic logic using the existential theory of invertible matrices. In this section, we consider diagrams with values in real vector spaces (or rational, but as we have seen in the previous section, both theories will coincide). We prove the decidability of the following two problems:\\
	\indent -- \textbf{bisimilarity:} given two finitary diagrams, are they bisimilar?\\
	\indent -- \textbf{diagram model-checking:} given a finitary diagram $F$, an object $c$ of its domain and a positive finitary state formula $S$, does $F,c \vDash S$ hold?

\subsection{Decidability of bisimilarity}

We start with the bisimilarity problem. Assume given two finitary diagrams $F$ and $G$, with domain $(\C, \leq)$ and $(\D, \preceq)$ respectively. The idea is to non-deterministically construct a bisimulation $R$, that is, a set of triples $(c,M,d)$ where $M$ is a matrix with coefficients in real (or rational) numbers satisfying the properties of a bisimulation from Section \ref{sec:bisim}. The only exception is that we will not guess explicitly the matrices $M$, but a formula in the existential theory of invertible matrices that encodes the fact that there exist some matrices $M$ such that the bisimulation constructed satisfies those properties.

\begin{algorithm}
\caption{Bisimilarity of finitary diagrams}
\begin{algorithmic}[1]
\Require Two finitary diagrams $\map{F}{\C}{\A}$ and $\map{G}{\D}{\A}$.
\Ensure Answer \textbf{Yes} iff $F$ and $G$ are bisimilar.
\State $S := \C\cup\D;\, R := \varnothing;\,lin := \varnothing;\,var := \varnothing$;
\While{$S$ is non empty} 
	\State Pick some $c \in S$. Let us assume that $c \in \C$, the other case is symmetric.
	\State Non-deterministically choose $d \in \D$ with $F(c) = G(d) = n$. 
	\State \textbf{if} $d$ does not exist \textbf{then FAIL end if};
	\State Create a fresh variable $X$ and add the pair $(X,n)$ to $var$;
	\State Add $(c,X,d)$ to $R$ and do not mark it;
	\While{there is a non-marked element in $R$}
		\State Pick a non-marked element $(c,X,d) \in R$, with $F(c) = G(d) = n$;
		\State Mark $(c,X,d)$;
		\State Non-deterministically choose a relation $$Q \subseteq \{(c',d') \mid (c < c' \wedge d \preceq d') \vee (c \leq c' \wedge d \prec d')\}$$ such that for every $c' > c$, there is $d' \succeq d$ with $(c',d')$ in $Q$, and symmetrically;
		\State $S := S\setminus (\{c'\mid c' \geq c\}\cup\{d'\mid d' \succeq d\})$
		\ForAll{$(c',d')$ in $Q$}
			\State Check if $F(c') = G(d') = m$, otherwise \textbf{FAIL};
			\State Create a fresh variable $X'$ and add the pair $(X',m)$ to $var$;
			\State Add $(c',X',d')$ to $R$ and do not mark it;
			\State Add the equation $G(d \preceq d').X = X'.F(c\leq c')$ to $lin$;
		\EndFor
	\EndWhile
\EndWhile
\State Let $\Phi$ be the formula of the theory of invertible matrices quantified by $\exists_n X$ for every $(X,n) \in var$ and whose predicates are the linear equations from $lin$.
\State\Return \textbf{YES} if $\Phi$ is valid, \textbf{FAIL} otherwise.
\end{algorithmic}
\label{alg:algo1}
\end{algorithm}

Consider the algorithm \ref{alg:algo1} written in pseudo-code. It maintains the bisimulation $R$ and two sets $var$, encoding the variables of the formula we are constructing and $lin$, encoding its predicates.

The algorithm always terminates. First, the innermost while loop terminates since after every loop an element $(c,X,d)$ is marked and only elements of the form $(c',X',d')$ with either $c < c'$ and $d \preceq d'$ or $c \leq c'$ and $d \prec d'$ are added. The outer loop terminates since after every loop at least one element of $S$ is removed.

Assume that there is an execution of the algorithm that answers \textbf{Yes}. Let $R$ and $\Phi$ constructed during this execution. Since the algorithm answers \textbf{Yes}, the formula $\Phi$ is satisfiable, that is, for every $(X,n) \in var$, there is an invertible matrix $M_X$ of size $n\times n$ such that for every equation $A.X=X'.B$ in $lin$, $A.M_X = M_{X'}.B$ holds. Let $R'$ be the set $\{(c,M_X,d) \mid (c,X,d) \in R\}$.
Then by construction of $R$ and $\Phi$, $R'$ is a bisimulation between $F$ and $G$.\\
Assume that there is a bisimulation $R'$ between $F$ and $G$. We show that there are non-deterministic choices that lead to the answer \textbf{Yes}. The idea is to ensure that every $(c,X,d)$ that belongs to $R$ at some point corresponds to an element $(c,f,d)$ of $R'$. To ensure this, we must:\\
	\indent 1. when choosing $d$ in line $7$, choose it such that there is $(c,f,d) \in R'$. It exists by definition of a bisimulation.\\
	\indent 2. when choosing $Q$ in line $17$, choose it in such a way that for every $(c',d') \in Q$, there is $(c',f',d')$ in $R'$ and that the element $(c,f,d) \in R'$ corresponding to $(c,X,d)$ satisfies that $G(d\leq d')\circ f = f'\circ F(c\leq c')$. Such a $Q$ always exists since $R'$ is a bisimulation.\\
With this, the algorithm does not \textbf{FAIL} and the formula $\Phi$ is valid: the assignment that map $X$ to the corresponding $f$ satisfies $\Phi$. Consequently, the algorithm answers \textbf{Yes}. Finally, this algorithm non-deterministically construct in exponential space a formula of exponential size in the size of the data. By Theorem 5, this algorithm is in NEXPSPACE. Consequently, by Savitch's theorem \cite{savitch70}, since NEXPSPACE = EXPSPACE:
\begin{theorem}
Knowing if two finitary diagrams are bisimilar in real or in rational numbers is decidable in EXPSPACE. 
\end{theorem}

\begin{example}
Consider the two finitary diagrams at the end of this Section, $F$ on the left, $G$ on the right. Let us apply a few steps of the algorithm on those two diagrams:
\begin{itemize}
	\item[1.] Pick $a$ and choose $0$. At this point $S = \{1,2,b,c,d\}$, $var = [(X_1,1)]$ and $R = [(0,X_1,a)]$ (we will only write the unmarked elements).
	\item[2.] Pick $(0,X_1,a)$ and choose $Q = \{(1,c), (2,d),(0,b)\}$. At this point, $S = \varnothing$, $var = [(X_1,1); (X_2,2);(X_3,1);(X_4,1)]$, $R = [(1,X_2,c) ; (2,X_3,d); (0,X_4,b)]$ and $lin = [\bigl(\begin{smallmatrix}0\\2\end{smallmatrix}\bigr).X_1 = X_2.\bigl(\begin{smallmatrix}1\\0\end{smallmatrix}\bigr) ; 6.X_1 = X_3 ; 2X_1 = X_4]$.
	\item[3.] Pick $(2, X_3, d)$ and choose $Q = \varnothing$. At this point, $R = [(1,X_2,c) ; (0,X_4,b)]$.
	\item[4.] Pick $(1,X_2,c)$ and choose $Q = \{(2,d)\}$. At this point, \[var = [(X_1,1); (X_2,2);(X_3,1);(X_4,1);(X_5,1)],\] $R = [(0,X_4,b), (2,X_5,d)]$ and \[lin = [\bigl(\begin{smallmatrix}0\\2\end{smallmatrix}\bigr).X_1 = X_2.\bigl(\begin{smallmatrix}1\\0\end{smallmatrix}\bigr) ; 6.X_1 = X_3 ; 2X_1 = X_4 ; \bigl(\begin{smallmatrix}4&3\end{smallmatrix}\bigr).X_2 = X_5.\bigl(\begin{smallmatrix}1&1\end{smallmatrix}\bigr)].\]
	\item[5.] $\ldots$
\end{itemize}
At the end, the algorithm produces \[var = [(X_1,1); (X_2,2);(X_3,1);(X_4,1);(X_5,1);(X_6,2);(X_7,1);(X_8,1)]\] and their linear equations: \begin{align*}lin = [&\bigl(\begin{smallmatrix}0\\2\end{smallmatrix}\bigr).X_1 = X_2.\bigl(\begin{smallmatrix}1\\0\end{smallmatrix}\bigr) ; 6.X_1 = X_3 ; 2X_1 = X_4 ; \bigl(\begin{smallmatrix}4&3\end{smallmatrix}\bigr).X_2 = X_5.\bigl(\begin{smallmatrix}1&1\end{smallmatrix}\bigr);\\ &\bigl(\begin{smallmatrix}0\\1\end{smallmatrix}\bigr).X_4 = X_6.\bigl(\begin{smallmatrix}1\\0\end{smallmatrix}\bigr);3.X_4 = X_7;\bigl(\begin{smallmatrix}4&3\end{smallmatrix}\bigr).X_6 = X_8.\bigl(\begin{smallmatrix}1&1\end{smallmatrix}\bigr)].\end{align*} The induced problem of invertible matrices is satisfiable, which means that both diagrams are bisimilar.
\begin{center}
	\begin{tikzpicture}[scale=1]
		\node (A) at (0,1) {$0$};
		\node (B) at (1,1) {$1$};
		\node (C) at (2,1) {$2$};
		\node (A') at (0,0) {$1$};
		\node (B') at (1,0) {$2$};
		\node (C') at (2,0) {$1$};
		\path[->,dotted]
		(A) edge (A')
		(B) edge (B')
		(C) edge (C');
		\path[->,font=\scriptsize]
		(A) edge (B)
		(A') edge node[below]{$\bigl(\begin{smallmatrix}1\\0\end{smallmatrix}\bigr)$} (B')
		(B) edge (C)
		(B') edge node[below]{$\bigl(\begin{smallmatrix}1&1\end{smallmatrix}\bigr)$} (C');
		\node (A) at (4,1) {$a$};
		\node (B) at (5,1) {$b$};
		\node (C) at (6,1) {$c$};
		\node (D) at (7,1) {$d$};
		\node (A') at (4,0) {$1$};
		\node (B') at (5,0) {$1$};
		\node (C') at (6,0) {$2$};
		\node (D') at (7,0) {$1$};
		\path[->,dotted]
		(A) edge (A')
		(B) edge (B')
		(C) edge (C')
		(D) edge (D');
		\path[->,font=\scriptsize]
		(A) edge (B)
		(A') edge node[below]{$\bigl(\begin{smallmatrix}2\end{smallmatrix}\bigr)$} (B')
		(B) edge (C)
		(B') edge node[below]{$\bigl(\begin{smallmatrix}0\\1\end{smallmatrix}\bigr)$} (C')
		(C) edge (D)
		(C') edge node[below]{$\bigl(\begin{smallmatrix}4&3\end{smallmatrix}\bigr)$} (D');
	\end{tikzpicture}
	\end{center}
\end{example}

\subsection{Decidability of the model checking}

Starting with a finitary diagram $F$, an element $c$ of its domain, and a positive finitary object formula $S$, we inductively construct two lists, initially empty, as previously:\\
	\indent -- $var$ of pairs $(X,n)$ where $X$ is a variable and $n$ an integer. This will stand for $\exists_n X$.\\
	\indent -- $lin$ of equations $A.X=Y.B$ where $X$ and $Y$ are variables and $A$ and $B$ are matrices.

The formula $S$ is of the form $[n]P$. We first check if $n = F(c)$. If it is not the case then we fail. Otherwise, let $X$ be a fresh variable. Add the pair $(X,n)$ to $var$. Continue with $F$, $c$, $X$, and $P$.

Now, assume that we consider the following data: a finitary diagram $F$, an element of its domain $c$, an $X$ with $(X,n)$ in $var$ for some interger $n$ and a positive finitary morphism formula $P$. Several cases:\\
	\indent -- if $P = ?S'$, continue with $F$, $c$ and $S'$,\\
	\indent -- if $P = \top$, stop,\\
	\indent -- if $P = P_1\wedge P_2$, first continue with $F$, $c$, $X$ and $P_1$. When this part terminates, continue with $F$, $c$, $X$ and $P_2$,\\
	\indent -- if $P = \langle M \rangle P'$, with $M$ of size $n_1\times n_2$. If $n_1 \neq F(c)$, then we fail. Otherwise, non-deterministically choose an element $c' \geq c$, with $F(c') = n_2$. If such a $c'$ does not exist, then we fail. Then, create a fresh variable $X'$, add $(X',n_2)$ to $var$ and $M.X = X'.F(c\leq c')$ to $lin$. Finally, continue with $F$, $c'$, $X'$ and $P'$.

If the algorithm does not fail, construct a formula $\Phi$ from $var$ and $lin$ as previously and check if it is satisfiable using the existential theory of invertible matrices. The formula $\Phi$ is non-deterministically constructed in polynomial time and so is of polynomial size. So, this algorithm is in NPSPACE and again, by Savitch's theorem \cite{savitch70}, since NPSPACE = PSPACE:
\begin{theorem}
Knowing if a finitary diagram satisfies a positive finitary formula (either in real or in rational numbers) is decidable in PSPACE.
\end{theorem}

\begin{example}
Let us consider the following positive finitary formula \[\phi = [1]\langle\bigl(\begin{smallmatrix}1\\0\end{smallmatrix}\bigr)\rangle\langle\bigl(\begin{smallmatrix}1&1\end{smallmatrix}\bigr)\rangle\top.\] It is not hard to check that $F,0 \vDash \phi$, and so that $G,a \vDash \phi$ (you can unroll the algorithm, the identities will give a solution of the problem of matrices). Let $H$ be the following diagram:
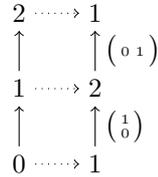
\begin{center}
	\begin{tikzpicture}[scale=1]
		\node (A) at (0,1) {$0$};
		\node (B) at (1,1) {$1$};
		\node (C) at (2,1) {$2$};
		\node (A') at (0,0) {$1$};
		\node (B') at (1,0) {$2$};
		\node (C') at (2,0) {$1$};
		\path[->,dotted]
		(A) edge (A')
		(B) edge (B')
		(C) edge (C');
		\path[->,font=\scriptsize]
		(A) edge (B)
		(A') edge node[below]{$\bigl(\begin{smallmatrix}1\\0\end{smallmatrix}\bigr)$} (B')
		(B) edge (C)
		(B') edge node[below]{$\bigl(\begin{smallmatrix}0&1\end{smallmatrix}\bigr)$} (C');
	\end{tikzpicture}
	\end{center}
	We will show that $H, 0 \not\vDash \phi$, and that $H$ is not bisimilar to $F$ and $G$. Let us unroll the algorithm on $H$, $0$ and $\phi$. We are in the first case, and we create a fresh variable $X_1$ and $var := [(X_1,1)]$. We then continue the algorithm with $H$, $0$, $X_1$ and $\langle\bigl(\begin{smallmatrix}1\\0\end{smallmatrix}\bigr)\rangle\langle\bigl(\begin{smallmatrix}1&1\end{smallmatrix}\bigr)\rangle\top$. We are then in the last case, and we can only choose $1$ without failing. So, $var = [(X_1,1);(X_2,2)]$ and $lin = [\bigl(\begin{smallmatrix}1\\0\end{smallmatrix}\bigr).X_1 = X_2.\bigl(\begin{smallmatrix}1\\0\end{smallmatrix}\bigr)]$. We continue with $H$, $1$, $X_2$ and $\langle\bigl(\begin{smallmatrix}1&1\end{smallmatrix}\bigr)\rangle\top$. We still are in the last case and we can only choose $2$ without failing. So, $var = [(X_1,1);(X_2,2);(X_3,1)]$ and $lin = [\bigl(\begin{smallmatrix}1\\0\end{smallmatrix}\bigr).X_1 = X_2.\bigl(\begin{smallmatrix}1\\0\end{smallmatrix}\bigr);\bigl(\begin{smallmatrix}1&1\end{smallmatrix}\bigr).X_2 = X_3.\bigl(\begin{smallmatrix}0&1\end{smallmatrix}\bigr)]$. Let us prove that we cannot solve this problem of invertible matrices. If we could, we would have that:
	$$X_1 = \bigl(\begin{smallmatrix}1&1\end{smallmatrix}\bigr).\bigl(\begin{smallmatrix}1\\0\end{smallmatrix}\bigr).X_1 = \bigl(\begin{smallmatrix}1&1\end{smallmatrix}\bigr).X_2.\bigl(\begin{smallmatrix}1\\0\end{smallmatrix}\bigr) = X_3.\bigl(\begin{smallmatrix}0&1\end{smallmatrix}\bigr).\bigl(\begin{smallmatrix}1\\0\end{smallmatrix}\bigr) = 0$$
	which is impossible since $X_1$ must be invertible.
\end{example}

%

\section{Future work}


As a future work, we would like to investigate the case of diagrams with values in $\mathbb{Z}$-modules (that is, Abelian groups), i.e., diagrams with values in matrices whose coefficients are integers, for which the existential theory is undecidable, but for which we can still decide some problems of matrices. Another interesting direction is the relation between our algorithm of Section~\ref{sec:algo} to find a bisimulation and the final chain algorithm \cite{adamek12}, which we let for a future work.

%
%
%
 \bibliographystyle{splncs04}
 \bibliography{refs}

\end{document}